\documentclass[12pt,preprint]{article}

\usepackage{amssymb}
\usepackage{amsmath}
\usepackage{graphics}
\usepackage{epsfig}
\renewcommand{\vec}[1]{{\bf #1}}
\newcommand{\eqb}{\begin{equation}}
\newcommand{\eqe}{\end{equation}}
\newcommand{\dmb}{\begin{displaymath}}
\newcommand{\dme}{\end{displaymath}}

\newcommand{\eab}{\begin{eqnarray}}
\newcommand{\eae}{\end{eqnarray}}

\newcommand{\e}{\mbox{e}}
\newcommand{\be}{\begin{equation}}
\newcommand{\ee}{\end{equation}}

\begin{document}
\begin{titlepage}
%\begin{flushright} 
%\end{flushright}

\begin{center}
\Large{Improved ground-state estimate by thermal resummation}\vspace{1.5cm}\\ 
\large{Carlos Falquez$^*$, Ralf Hofmann$^{**}$, and Tilo Baumbach$^*$}
\end{center}
\vspace{2.0cm} 
\begin{center}
{\em $\mbox{}^{*}$ 
Laboratorium f\"ur Applikationen der Synchrotronstrahlung (LAS)\\ 
Karlsruher Institut f\"ur Technologie (KIT)\\ 
Postfach 6980\\ 
76128 Karlsruhe, Germany}\vspace{0.5cm}\\ 
{\em $\mbox{}^{**}$ Institut f\"ur Theoretische Physik\\ 
Universit\"at Heidelberg\\ 
Philosophenweg 16\\ 
69120 Heidelberg, Germany}
\end{center}
\vspace{1.0cm}
\begin{abstract}
For the deconfining phase of SU(2) Yang-Mills thermodynamics and for high temperatures 
we point out that a linear dependence on temperature of a one-loop selfconsistently resummed 
thermal correction to the pressure and the energy density takes place despite a 
quartic dependence arising from an unsummed two-loop correction. This linearity is hierarchically 
smaller than the one belonging to the tree-level estimate of 
the thermal ground-state. We discuss and interpret this result.       
\end{abstract} 
\end{titlepage}

\noindent{\sl Introduction.} In addressing the deconfining (and preconfining) thermodynamics of an 
SU(2) Yang-Mills theory the notion of a thermal ground state proves useful 
\cite{HerbstHofmann2004,Hofmann2005,Hofmann2007}. Composed of 
interacting calorons and anticalorons of topological charge modulus 
$|Q|=1$ \cite{Nahm,vanBaal,LeeLu} an a priori estimate for this ground state emerges upon a spatial 
coarse-graining over noninteracting (anti)calorons of trivial holonomy \cite{HarringtonShepard1977} down to a 
a certain resolution. The latter selfconsistently and in dependence of temperature and an integration constant 
is set by the emerging, spacetime-homogeneous 
modulus $|\phi|$ of an adjoint, nonpropagating scalar field $\phi$. 

The entirety of hard quantum 
fluctuations (higher resolving power than $|\phi|$) associated with fundamental field configurations of 
trivial topology, whose influence on (anti)calorons is 
void of full analytical access beyond the semiclassical approximation \cite{Diakonov2004}, 
is encoded in terms of 
a pure-gauge configuration $a_\mu^{\tiny\mbox{gs}}$ after 
coarse-graining. The field configuration $a_\mu^{\tiny\mbox{gs}}$ introduces finite ground-state pressure and energy 
density thus lifting the pure BPS situation described by $\phi$ alone.  
Conceptually, there is some resemblence with the treatment of 
quantum fluctions in perturbation theory where the subtraction of 
infinities by counter terms, respecting the form of the classical 
Yang-Mills action, consistently ignores the influence of ultraviolet physics 
at a given resolution and at an arbitrary loop order 
\cite{tHooftVeltman,Zinn-Justin}. We assume the existence 
of the Yang-Mills partition function. The latter could be formulated 
as a weighted average over fundamental fields at a high resolution. The according weight should then be derivable from 
the action of the infinitely resolved system 
in the sense of a Borel resummable, approximating series obtained by 
consistent order-by-order removal of (or an all-order average over) the ultraviolet 
physics. Then the effective action at resolution 
$|\phi|$ neatly splits into the classical Yang-Mills term describing the propagation and 
interactions among topologically trivial fluctuations, an interaction term 
between $Q=0$ and $|Q|=1$ configurations given by the square of the covariant 
derivative acting on $\phi$, $(D_\mu\phi)^2$ (Higgs mechanism for two out of three 
propagating directions in the SU(2) algebra), and a 
potential $V$ for the field $\phi$ \cite{Hofmann2005}. It is conceivable that a formulation of the  
partition function at a higher resolution than $|\phi|$ exhibits that 
the interaction between topological and field configurations with 
$Q=0$ effectively cuts off the number of higher dimensional operators \cite{PFA1,PFA2,PFA3,PFA4} 
in the corresponding effective action, limits the number of 
irreducible $Q=0$ loops, and poses a limit to the topological 
charge modulus $|Q|$ of relevant field configurations. The usefulness of the effective theory 
\cite{HerbstHofmann2004,Hofmann2005,Hofmann2007} at resolution $|\phi|$ is 
linked to the fact that this is explicitly demonstrable.    

In the effective theory \cite{HerbstHofmann2004,Hofmann2005,Hofmann2007} 
the temperature dependence of the gauge coupling 
$e$ is dictated by the Legendre transformation between the pressure and 
energy density for tree-level quasiparticles (Higgs mechanism) fluctuating freely above 
the estimate of the thermal ground state expressed by $\phi$, its potential $V$, and 
$a_\mu^{\tiny\mbox{gs}}$. At high temperature one 
has $e\equiv \sqrt{8\pi}$. A direct contribution to the linear temperature dependence of the 
pressure $P^{\tiny\mbox{gs}}$ and the energy density $\rho^{\tiny\mbox{gs}}$ of the 
thermal ground state arises via the potential 
$V(\phi)=\mbox{tr}\,\frac{\Lambda^6}{\phi^2}=4\pi\Lambda^3T$. Thus 
%********
\eqb
\label{apestgs}
\frac{\rho^{\tiny\mbox{gs,V}}}{T^4}=-\frac{P^{\tiny\mbox{gs,V}}}{T^4}=2(2\pi)^4\,\lambda^{-3}\sim 3117.09\,
\lambda^{-3}\,,
\eqe
%********* 
where $\lambda\equiv\frac{2\pi T}{\Lambda}$, and the Yang-Mills scale $\Lambda$ is related to 
the critical temperature $T_c$ of the deconfining-preconfining 
transition as $\Lambda=\frac{2\pi}{13.87}\,T_c$. At high temperature there are in addition 
linear contributions $\Delta P^{\tiny\mbox{gs,1-loop}}$ and 
$\Delta\rho^{\tiny\mbox{gs,1-loop}}$ to $P^{\tiny\mbox{gs}}$ and $\rho^{\tiny\mbox{gs}}$, respectively, 
arising from the free fluctuations of tree-level massive modes \cite{GiacosaHofmann2007}. 
Namely, one has
%************
\eqb
\label{fluctregs}
\frac{\Delta\rho^{\tiny\mbox{gs,1-loop}}}{T^4}=\frac{\Delta P^{\tiny\mbox{gs,1-loop}}}{T^4}=
-\frac14\,a^2=-2(2\pi)^4\,\lambda^{-3}\sim -3117.09^4\,\lambda^{-3}\,,
\eqe
%*************
where $a\equiv\frac{2e|\phi|}{T}=\frac{8\sqrt{2}\pi^2}{\lambda^{3/2}}$. 
Notice that by virtue of the linear dependence on temperature 
of $P^{\tiny\mbox{gs,V}}+\Delta P^{\tiny\mbox{gs,1-loop}}$ the Legendre transformation yields 
$\rho^{\tiny\mbox{gs,V}}+\Delta\rho^{\tiny\mbox{gs,1-loop}}=0$ which, indeed, is the case, 
compare Eqs.\,(\ref{apestgs}) and (\ref{fluctregs}). 

The main purpose of this report is to demonstrate that at high temperature 
the ground-state inherent linear $T$ 
dependence in certain contributions to the total energy density and the total pressure 
is also generated by the selfconsistent one-loop propagation of 
the massless modes albeit subject to a hierarchically smaller 
coefficient compared to that of Eqs.\,(\ref{apestgs}) and (\ref{fluctregs}). 

Recalling that there is a leading quartic dependence on high 
temperature \cite{SchwarzHofmannGiacosa2007,KellerLudescherHofmannGiacosa2008} 
of the two-loop correction to the pressure, $\Delta P^{\tiny\mbox{2-loop}}$, 
this may come as a surprise. Specifically, 
one has \cite{SchwarzHofmannGiacosa2007,KellerLudescherHofmannGiacosa2008}
%******
\eqb
\label{2-loop}
\Delta P^{\tiny\mbox{2-loop}}=-\frac{4\pi^2}{45}\times 4.39\times 10^{-4}\, T^4\,
\eqe
%****** 
for $T\gg T_c$. While the correction 
$\Delta P^{\tiny\mbox{2-loop}}$ of Eq.\,(\ref{2-loop}) is interpreted 
as the loss in pressure of a thermal gas of tree-level 
massless, effective gauge modes due to the emergence of 
large-holonomy calorons and their 
subsequent dissociation into screened and 
long-lived monopole-antimonopole pairs \cite{KellerLudescherHofmannGiacosa2008} 
the above mentioned linear dependence, originating from a selfconsistent resummation of the one-loop irreducible 
insertion of the one-loop polarization tensor, represents a positive 
contribution to the pressure. That is, a {\sl radiatively} 
induced effect with ground-state characteristics (linear $T$ dependence) 
emerges as a result of the selfconsistent propagation 
of tree-level massless modes through 
a sea of stable and screened monopole-antimonopole pairs. 

\begin{figure}
\begin{center}
\leavevmode
\leavevmode
%\epsffile[80 25 534 344]{}
\vspace{4.3cm}
\includegraphics{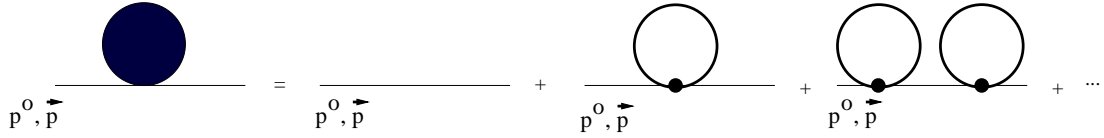}
\end{center}
\caption{The Dyson series for the resummation of the one-loop 
irreducible contribution to the polarization tensor of the massless mode. A 
thin line refers to the propagation of the tree-level massless, 
a thick line to the propagation of tree-level massive modes. \label{Fig-1}}      
\end{figure}

\begin{figure}
\begin{center}
\leavevmode
\leavevmode
%\epsffile[80 25 534 344]{}
\vspace{4.3cm}
\includegraphics{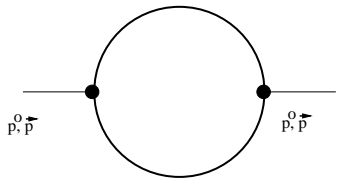}
\end{center}
\caption{Potential one-loop contribution to the screening function $G$. 
On the `mass' shell, generated by the series in Fig.\,\ref{Fig-1}, this contribution, 
however, vanishes \cite{LudescherHofmann2008}. \label{Fig-2}}      
\end{figure}

\begin{figure}
\begin{center}
\leavevmode
\leavevmode
%\epsffile[80 25 534 344]{}
\vspace{6.3cm}
\includegraphics{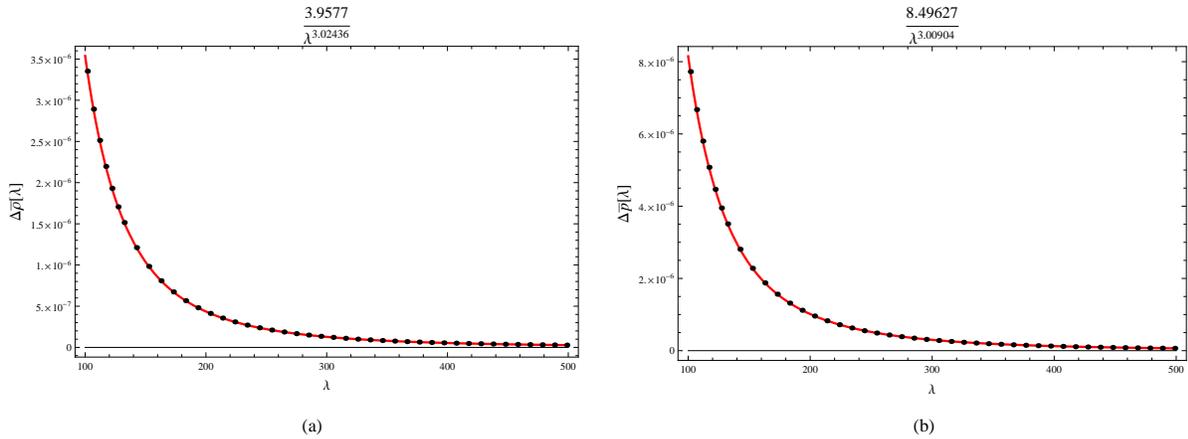}
\end{center}
\caption{The high-temperature $\lambda$ dependence of (a) $\Delta\bar{\rho}$ and (b) 
$\Delta\bar{P}$. Dots correspond to computed values, the line is a fit to this data. \label{Fig-3}}      
\end{figure}

\noindent{\sl Massless quasiparticles after resummation.} 
In Fig.\,\ref{Fig-1} 
the diagrammatic basis for the resummation of 
the one-loop contribution to the on-shell polarization 
tensor of the massless mode is shown. This resummation into a selfconsistently modified dispersion law for 
transverse, propagating, tree-level massless 
modes was performed in \cite{LudescherHofmann2008} and is 
summarized in terms of the screening function $G(T,|\vec{p}|)$: 
%*******
\eqb
\label{displa}
p_0^2=\vec{p}^2+G(T,|\vec{p}|)\ \Leftrightarrow\ Y^2=X^2+\frac{G}{T^2}(\lambda,X)\,,
\eqe
%*******
where $p_0$ is the mode's energy, $\vec{p}$ its spatial momentum, 
and $Y\equiv\frac{p_0}{T}$, $X\equiv\frac{|\vec{p}|}{T}$. Details on the properties of $G$ are presented in 
\cite{LudescherHofmann2008}, here we mention only that $G$ is positive for low 
momenta $|\vec{p}|$ (screening) and that it turns negative above some critical, temperature dependent 
momentum (antiscreening). For 
increasingly large $|\vec{p}|$ there is an exponentially fast decay of $|G|$. 
On the 'mass' shell 
of Eq.\,(\ref{displa}) there is no one-loop irreducible, imaginary contribution to the polarization 
tensor arising from the diagram depicted in Fig.\,\ref{Fig-2}.       
%***********************
%************************
Resummation of the polarization tensor into a contribution 
$P^{\tiny\mbox{1-loop,res}}$ to the total pressure arising from 
free but radiatively shifted tree-level massless quasiparticles 
is performed by connecting the external legs on the left-hand side 
of Fig.\,\ref{Fig-1}. For dimensionless pressure and energy density one has:
%*********
\eab
\label{QPPEN}
\frac{P^{\tiny\mbox{1-loop,res}}}{T^4}&=&\frac{1}{\pi^2}\int_0^\infty dX\,\frac{X^2\sqrt{X^2+G/T^2}}{\e^{\sqrt{X^2+G/T^2}}-1}\,,\nonumber\\ 
\frac{\rho^{\tiny\mbox{1-loop,res}}}{T^4}&=&-\frac{1}{\pi^2}\int_0^\infty dX\,X^2\log\,(1-\e^{-\sqrt{X^2+G/T^2}})\,.
\eae
%***********************
%************************
We define the dimensionless corrections $\frac{\Delta P^{\tiny\mbox{1-loop,res}}}{T^4}$ and 
$\frac{\Delta P^{\tiny\mbox{1-loop,res}}}{T^4}$ to the contributions of 
free, massless particles as 
%******
\eqb
\label{defcorr}
\frac{\Delta P^{\tiny\mbox{1-loop,res}}}{T^4}\equiv
\Delta\bar{P}\equiv\frac{P^{\tiny\mbox{1-loop,res}}}{T^4}-\frac{\pi^2}{45}\,,\ \ \ \ \ 
\frac{\Delta \rho^{\tiny\mbox{1-loop,res}}}{T^4}\equiv\Delta\bar{\rho}\equiv\frac{\rho^{\tiny\mbox{1-loop,res}}}{T^4}-\frac{\pi^2}{15}\,.
\eqe
%*******
In Fig.\,\ref{Fig-3} both $\Delta\bar{P}$ and $\Delta\bar{\rho}$ 
are depicted as a function of $\lambda$ within the high-temperature interval 
$100\le\lambda\le 500$ (recall that $\lambda_c=13.87$). An excellent fit 
to the following power dependences 
%*************
\eqb
\label{defcorrA}
\Delta\bar{P}\equiv c_P\lambda^{\delta}\,,\ \ \ \ 
\Delta\bar{\rho}\equiv c_\rho\lambda^{\gamma}\,
\eqe
%**************** 
reveals that
%********
\eqb
\label{fitvalues}
c_P=8.49627\,,\ \ \ \delta=-3.00904\,,\ \ \ c_\rho=3.9577\,,\ \ \ \gamma=-3.02436\,. 
\eqe
%*********
Eq.\,(\ref{fitvalues}) represents the main result of our present work: 
Through infinite and selfconsistent one-loop 
resummation the power and the sign in front of this power in temperature 
of a fixed-order correction to pressure and energy density is profoundly changed!  
Specifically, by resumming the selfconsistent one-loop polarization of the massless modes, 
we generate a linear correction which is about three orders of magnitude smaller than the a priori 
estimate given by $P^{\tiny\mbox{gs,V}}$ and $\rho^{\tiny\mbox{gs,V}}$, 
compare Eqs.\,(\ref{apestgs}), (\ref{defcorrA}), 
and (\ref{fitvalues}). That this correction is linear 
to a very good approximation is not a chance-result: Selfconsistent 
propagation, which is indirectly influenced by the a priori estimate 
of the thermal ground state through the quasiparticle mass of tree-level heavy modes, 
yields a correction improving this very estimate. 
In calculating the contribution of higher irreducible 
loop orders to the polarization tensor we observe a hierarchic decrease 
and expect a termination at a finite order \cite{Hofmann2006,KavianiHofmann2007} such 
that a consideration of those contributions in the resummation process practically does not change 
our result obtained here by one-loop resummation. 

\begin{figure}
\begin{center}
\leavevmode
\leavevmode
%\epsffile[80 25 534 344]{}
\vspace{5.3cm}
\includegraphics{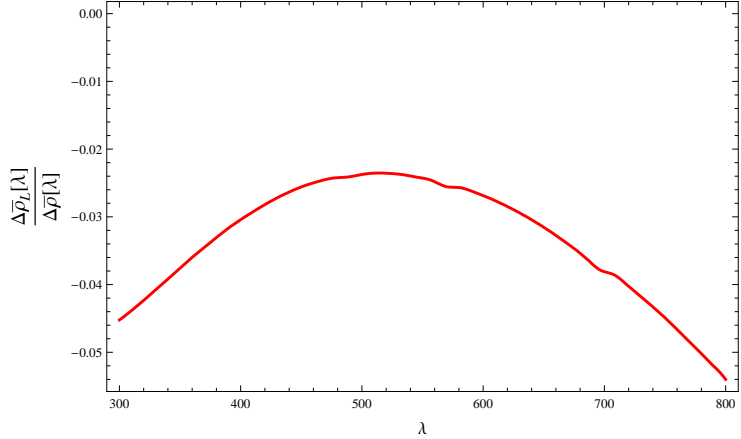}
\end{center}
\caption{The $\lambda$ dependence of the ratio $\frac{\Delta\bar{\rho}_L}{\Delta\bar{\rho}}$, where 
$\Delta\rho_L^{\tiny\mbox{1-loop,res}}\equiv T^4\,\Delta\bar{\rho}_L$ is defined in Eq.\,(\ref{leg}), and 
$300\le\lambda\le 800$. The figure clearly expresses that the corrections to 
pressure and energy density of a thermal gas of 
massless particles due to one-loop resummation are thermodynamically not 
selfconsistent because radiative corrections shifting the dispersion law of tree-level massive modes 
are not considered.\label{Fig-4}}      
\end{figure}

Corrections $\Delta P^{\tiny\mbox{1-loop,res}}$ and $\Delta\rho^{\tiny\mbox{1-loop,res}}$ 
are not thermodynamically 
selfconsistent. Namely, defining $\Delta\rho_L^{\tiny\mbox{1-loop,res}}$ 
to be the Legendre transformation 
of $\Delta P^{\tiny\mbox{1-loop,res}}$,
%*********
\eqb
\label{leg}
\Delta\rho_L^{\tiny\mbox{1-loop,res}}\equiv 
T\frac{d\Delta P^{\tiny\mbox{1-loop,res}}}{dT}-\Delta P^{\tiny\mbox{1-loop,res}}\,,
\eqe
%**********
we observe from Fig.\,\ref{Fig-4} that by no means 
$\Delta\rho_L^{\tiny\mbox{1-loop,res}}=\Delta\rho^{\tiny\mbox{1-loop,res}}$ 
(or $\Delta\bar{\rho}_L=\Delta\bar{\rho}$); even their 
signs are different. But in contrast to the correction $\Delta P^{\tiny\mbox{2-loop}}$ of 
Eq.\,(\ref{2-loop}), which is by far dominating the two-loop correction to the pressure 
at high temperatures and thus needs to be thermodynamically selfconsistent by 
itself no 
such relation exists between $\Delta P^{\tiny\mbox{1-loop,res}}$ and 
$\Delta\rho^{\tiny\mbox{1-loop,res}}$. On the contrary, thermodynamical selfconsistency after one-loop 
resummation is only expected to occur when corrections due to a shift in 
dispersion law of the tree-level massive modes are taken into account at resummed one-loop 
level. Small deviations from thermodynamical selfconsistency after taking these contributions 
into account arise from the use of tree-level quasiparticle consistency at a higher radiative 
order. Since the contributions to the pressure arising from radiative corrections 
follow a large hierarchy \cite{Hofmann2006,KavianiHofmann2007} we are assured that the 
demand for thermodynamical selfconsistency at a higher loop order introduces small 
changes to the evolution of $e$ which in turn induce very small 
changes to the radiative corrections themselves. For practical purposes it thus suffices to 
work at tree-level selfconsistency to judge the selfconsistency of 
radiative corrections. The shift in disperion law for tree-level massive modes 
is harder to compute than for tree-level massive modes since: 
(i) The resummation invokes two separate one-loop 
contributions to the polarization tensor arising from a massive and a massless 
tadpole. (ii) The polarization tensor is not 
transverse for propagating massive modes, and thus more than just a single 
screening function need to be considered. We leave this to future investigation.   
%***********************
%************************
%***********************
%************************

\noindent{\sl Summary and conclusions.} To summarize, we have shown 
numerically that for high temperatures the selfconsistent resummation of 
the one-loop polarization tensor of the tree-level 
massless mode generates a linear dependence on temperature in the according 
correction to the pressure and the energy 
density of the thermal gas of massless particles. This is remarkable 
because a two-loop correction to the pressure, which dominates all fixed-order 
radiative corrections at high-temperature, depends quartically on 
temperature. While this fixed-order correction is interpreted as an investment of energy 
by massless gauge modes into the emergence of screened but stable monopole-antimonopole pairs upon strong 
(anti)calorons deformation (large temporary holonomy) 
and subsequent dissociation \cite{Diakonov2004}, the correction due to 
selfconsistent resummation describes the propagation in a 
preexisting (yet unresolvable \cite{KellerLudescherHofmannGiacosa2008}) 
sea of stable, screened monopole-antimonopole pairs. A linear dependence on 
temperature of the pressure and energy-density correction 
is suggestive for the ground-state estimate \cite{Hofmann2005,Hofmann2007,GiacosaHofmann2007} 
being improved by a radiative dressing of tree-level 
massless modes.

\end{document}